\documentclass[letterpaper, 10pt]{article}
\usepackage[utf8]{inputenc}
\usepackage[top=1in, bottom=1in, left=1in, right=1in]{geometry}
\usepackage{float}
\usepackage{color}
\usepackage{tikz}

\usepackage{titling}
\RequirePackage[explicit]{titlesec}
\titlespacing*{\section}{0pt}{0.5em}{0.3pt}
\titlespacing*{\subsection}{0pt}{0.35em}{0pt}
\titlespacing*{\subsubsection}{0pt}{0.25em}{0pt}

\usepackage{caption}
\RequirePackage{fancyhdr}
\usepackage{lastpage}
\usepackage{empheq}

\usepackage{mathpazo}
\usepackage{xcolor}
\definecolor{GT}{RGB}{0, 0, 0}
\definecolor{Conv}{RGB}{247,37,133}
\definecolor{MC}{RGB}{114,9,183}
\definecolor{MVN}{RGB}{67,97,238}
\definecolor{FE}{RGB}{76,201,240}

\usepackage[
  colorlinks=true,
  linkcolor=blue,
  anchorcolor=blue,
  citecolor=blue,
  filecolor=blue,
  menucolor=blue,
  runcolor=blue,
  urlcolor=blue]{hyperref}
\usepackage{amsmath, amsthm, amssymb, amsfonts}
\usepackage{upgreek}
\usepackage{dsfont}

\usepackage{enumerate}
\usepackage{enumitem}

\usepackage{graphicx}
\usepackage{subcaption}

\usepackage{multirow}

\usepackage[numbers, sort&compress]{natbib}

\title{Forward and Inverse Modeling of Depth-of-Field Effects in Background-Oriented Schlieren}

\author{
  Joseph P. Molnar$^1$,
  Elijah J. LaLonde$^2$,
  Christopher S. Combs$^2$,
  Olivier L{\'e}on$^3$,\\
  David Donjat$^3$, and
  Samuel J. Grauer$^{1,}$\thanks{Corresponding author: \href{mailto:sgrauer@psu.edu}{sgrauer@psu.edu}}\vspace*{.3em}\\
  {\small $^1$Department of Mechanical Engineering, Pennsylvania State University}\vspace*{-.25em}\\
  {\small $^2$Department of Mechanical Engineering, University of Texas at San Antonio}\vspace*{-.25em}\\
  {\small $^3$D{\'e}partement Multi-Physique pour l'{\'E}nerg{\'e}tique, Office National d'{\'E}tudes et de Recherches A{\'e}rospatiales}\vspace{-1em}}
\date{}

\begin{document}

\maketitle
\setcounter{footnote}{3}
\vspace*{-2em}

\begin{abstract}
\noindent We report a novel ``cone-ray'' model of background-oriented schlieren (BOS) imaging that accounts for depth-of-field effects. Reconstructions of the density field performed with this model are far more robust to the blur associated with a finite aperture than conventional reconstructions, which presume a ``thin-ray'' pinhole camera. Our model is characterized and validated using forward evaluations based on simulated and experimental BOS measurements of buoyancy-driven flow and hypersonic flow over a sphere. Moreover, we embed the model in a neural reconstruction algorithm, which is demonstrated with a total variation penalty as well as the compressible Euler equations. Our cone-ray technique dramatically improves the accuracy of BOS reconstructions: the shock interface is well-resolved in all our tests, irrespective of the camera's aperture setting, which spans $f$-numbers from 22 down to 4.\par\vspace{.5em}

\end{abstract}
\vspace*{2em}

\section{Introduction}
\label{sec:introduction}
Schlieren imaging has long been employed to visualize density gradients in high-speed, buoyancy-driven, and thermally-driven flows \cite{Settles2001}. Although schlieren is mostly used as a qualitative or semi-quantitative tool,\footnote{``Semi-quantitative'' applications include locating flow features like separation lines, measuring shock angles, and the like.} there are several fully-quantitative variants, including calibrated knife-edge or grid-type schlieren \cite{Settles2017}, rainbow schlieren \cite{Howes1984, Greenberg1995}, and background-oriented schlieren (BOS) \cite{Raffel2015}: see \cite[Ch.~10]{Settles2001} for a comprehensive overview of these methods. Compared to other schlieren systems, BOS setups are cheap and easy to assemble, they can acquire large-scale measurements in a harsh environment, and BOS inherently delivers a deflection \textit{vector field} (instead of just one component normal to the knife edge or cutoff) \cite{Elsinga2004, Hargather2012, Mariani2019}. Consequently, BOS has been used to characterize a diverse set of target flows, including compressible jets \cite{vanHinsberg2014, Nicolas2017a, Nicolas2017b, Lee2021}, external \cite{Venkatakrishnan2004, Ota2011, Sourgen2012, Davami2023} and internal \cite{Bustard2023} shock formations, blade wakes \cite{Alhaj2010} and blade-tip vortices \cite{Kindler2007, Bauknecht2014, Raffel2014}, flames \cite{Grauer2018, Liu2020, Choudhury2022}, sprays \cite{Lee2012}, boundary layers \cite{Zhu2023}, and blast waves \cite{Mizukaki2014, Winter2019, Gomez2022}, to name a few. This paper presents a new measurement model for BOS which accounts for depth-of-field effects, thereby increasing the accuracy of forward and inverse computations, i.e., simulating BOS data and reconstructing flow states, respectively.\par

\begin{figure*}[ht]
    \centering
    \includegraphics[width=5in]{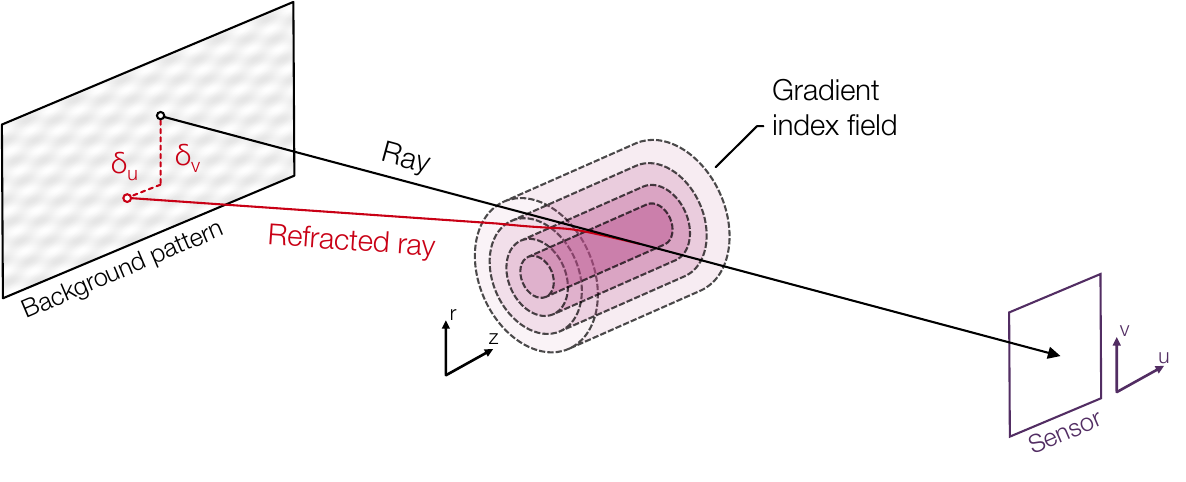}
    \caption{Schematic of BOS with an exaggerated deflection.}
    \label{fig:setup}
\end{figure*}

In BOS, a camera is focused through the working fluid onto a background pattern. When the flow is active, refraction within the fluid distorts images of said pattern. A pair of reference (flow-off) and active (flow-on) images are processed with a computer vision algorithm to estimate the deflection of light at each pixel. Deflections may be processed with a tomography algorithm to recover the underlying density field. A typical BOS setup is illustrated in Fig.~\ref{fig:setup}. Unfortunately, since the camera is focused on the background plate, the flow itself is out of focus, leading to blurry deflections which obscure features of interest in the flow \cite{Nicolas2017a}. At the same time, it is often desirable to open the camera's aperture: to collect more light and thereby increase the signal-to-noise ratio (SNR), to reduce the exposure time to better ``freeze'' the flow, or some combination of these goals. Of course, opening the aperture shrinks the camera's depth of field and worsens blur. Addressing adverse ``depth-of-field effects'' in BOS is thus a key research challenge.\footnote{Depth-of-field effects can be ameliorated with a telecentric imaging setup \cite{Zhou2023b}, but doing so comes at the cost of a conventional schlieren lens system and reduces the achievable field of view.}\par

Tomographic reconstruction entails the inversion of a forward measurement model, and the accuracy of reconstructions is contingent upon the accuracy of this model, that is, the degree to which it successfully predicts measurements of a known flow \cite{Kaipio2006}. To date, to the best of our knowledge, all BOS measurement models used for tomography have presumed an idealized pinhole camera configuration \cite{Raffel2015, Grauer2020}.\footnote{It is common to augment the pinhole camera with a lens distortion model, but the use of ``thin rays'' in BOS tomography is universal.} A pinhole camera records light that passes through an infinitesimal aperture, which corresponds to ``thin,'' 1D rays. In reality, camera apertures accept a cone of light that is focused down to a small spot on the detector. The cone has finite width in the measurement volume, resulting in blurry measurements associated with the camera's limited depth of field. Modeling conical rays as thin lines thus leads to a systematic overprediction of the peak deflection in forward calculations, and this model discrepancy is a source of reconstruction errors in the inverse procedure.\par

Various forward imaging models have been assessed for other tomographic modalities, including X-ray radiography, particle image velocimetry (PIV), and emission imaging. For instance, there is a range of X-ray source and detector configurations, which typically correspond to a parallel-, fan-, or cone-beam model \cite{Withers2021}. Due to the large number (hundreds to thousands) of view angles used in many of these systems, geometric simplifications are employed to model radiography at a low computational cost \cite{Long2013}. Further, while blur can arise when the source aperture or detector pixels are large, it is generally neglected in X-ray imaging. PIV features light scattering off quasi-point particles. Images of the scattered light are often simulated using an empirical ``optical transfer function'' that accounts for the scattering phase function as well as blur, diffraction, and astigmatism effects \cite{Schanz2012, Zigunov2023}. This approach is viable for point particles, but calibrating an empirical transfer function for imaging of a continuous field (emission tomography) or through it (BOS) is much more challenging. Walsh et al. \cite{Walsh2000} presented the first systematic analysis of errors associated with thin-ray -- viz., pinhole -- imaging models used in 2D chemiluminescence tomography. The authors found that neglecting the beam geometry caused reconstruction errors up to 30\%. Floyd \cite{Floyd2009} performed a similar analysis for 3D reconstructions and introduced the use of cylindrical- and conical-beam models. More recently, the group of Cai developed a comprehensive imaging model for volumetric emissions from chemiluminescence, incandescence, and fluorescence by extending the concept of a point spread function to voxels and linear finite elements \cite{Yu2017, Liu2021}. In the context of BOS, Rajendran et al. \cite{Rajendran2019a} performed a \textit{forward} Monte Carlo simulation of BOS imaging that inherently included depth-of-field effects. Kirmse, Amjad, and coworkers \cite{Kirmse2011, Amjad2020} also attempted to model blur in forward simulations of BOS by convolving a blur kernel with the density field from a computational fluid dynamics (CFD) simulation. Amjad et al. noted that blurry data produced significant artifacts in reconstructions of a smooth jet flow, increasing errors in the refractive index field by up to 35\% (similar to the findings of Walsh). However, there has been no attempt to correct or circumvent these errors in BOS until now.\par

This paper analyzes depth-of-field effects in BOS through two representative flow cases. First, we study BOS using a numerical buoyancy-driven turbulence data set, which features a wide range of scales and large deflections. Second, we perform numerical and experimental tests using hypersonic flow over a spherical bluff body. The latter case is a canonical flow with a bow shock whose location and density ratio have been well characterized \cite{VanDyke1958, VanDyke1959}. In what follows, we present the working principles of BOS, propose an imaging model that includes depth-of-field effects, and describe our reconstruction framework. Thereafter, Sec.~\ref{sec:cases} outlines the flow cases considered in this work, and Sec.~\ref{sec:modeling} reports an assessment of our model. We analyze depth-of-field effects in BOS using both forward and inverse problems. This paper conclusively demonstrates the need for and utility of an aperture-corrected forward model in BOS tomography.\par

\section{Background-Oriented Schlieren}
\label{sec:BOS}
Tomographic BOS can be divided into three inverse problems: deflection sensing, reconstruction, and solving a Poisson equation. First comes deflection sensing, but the latter steps are reversible and may be combined: either in Fourier \cite{Goldhahn2007} or physical \cite{Nicolas2016} space. It is also possible to perform all three steps at once in ``unified BOS'' (UBOS) \cite{Grauer2020}. A graphical overview of these pathways can be found in Fig.~1 of \cite{Molnar2023a}. This section presents optical flow for deflection sensing and a continuous model of refraction, with special attention paid to the ray geometry in the latter subsection. After that, we briefly review a neural reconstruction technique for BOS tomography, which we employ for inverse analysis in this paper.\par

\subsection{Optical Flow}
\label{sec:BOS:optical flow}
Optical flow is the prevailing method for determining the magnitude and orientation of deflections from a pair of reference and distorted images \cite{Atcheson2009, Schmidt2021, Cakir2023}. This technique assumes that the brightness of features in the background pattern is conserved across each image pair \cite{Davies2004}, which is formalized as follows:
\begin{equation}
    I\mathopen{}\left(\mathbf{u}, t\right) = I\mathopen{}\left(\mathbf{u} + \boldsymbol\updelta, t + \Delta t\right).
    \label{equ:intensity constancy}
\end{equation}
Here, $I$ is the intensity of an image at the sensor position $\mathbf{u} = [u, v]^\top$ and time $t$. The left side of Eq.~\eqref{equ:intensity constancy} corresponds to the reference image while the right side describes an intensity field distorted by deflections, $\boldsymbol\updelta = [\delta_\mathrm{u}, \delta_\mathrm{v}]^\top$, effectively warping the reference image. In principle, the intensity and deflection fields are continuous; in practice, they are discrete and generally resolved at the pixel centroids. The time between images, $\Delta t$, is arbitrary in BOS and may be set to unity for convenience. Lastly, in the context of BOS, it is commonly assumed that changes in the scene are produced by refraction, alone, as opposed to motion, shadows, reflections, and the like. Hence, variable illumination, facility vibrations, and light scattering off vapor are common sources of error in estimates of $\boldsymbol\updelta$.\par

When deflections are sufficiently small, Eq.~\eqref{equ:intensity constancy} can be approximated with a first-order Taylor expansion, 
\begin{subequations}
    \label{equ:Taylor series}
    \begin{align}
        I &\approx I + \frac{\mathrm{d}I}{\mathrm{d}t} \,\Delta t \\
        &= I + \left(
        \frac{\partial I}{\partial u} \frac{\partial u}{\partial t} +
        \frac{\partial I}{\partial v} \frac{\partial v}{\partial t} +
        \frac{\partial I}{\partial t}\right)\Delta t \\
        &= I
        + \frac{\partial I}{\partial u} \delta_\mathrm{u}
        + \frac{\partial I}{\partial v} \delta_\mathrm{v} 
        + \frac{\partial I}{\partial t} \Delta t,
    \end{align}
\end{subequations}
evaluated at $(\mathbf{u},t)$. Setting $\Delta t$ to one in Eq.~\eqref{equ:Taylor series} yields the linear optical flow equation,
\begin{equation}
    \boldsymbol\updelta \cdot \nabla I = -\frac{\partial I}{\partial t},
    \label{equ:linear optical flow}
\end{equation}
where $\nabla$ is the gradient operator in image space. Writing out Eq.~\eqref{equ:intensity constancy} or \eqref{equ:linear optical flow} for each pixel produces an underdetermined system of equations having two unknowns per equation: $\delta_\mathrm{u}$ and $\delta_\mathrm{v}$. Numerous closures have been developed for optical flow, the most popular of which are the Horn--Schunck \cite{Horn1981} and Lucas--Kanade \cite{Lucas1981} variants. Equation~\eqref{equ:linear optical flow} can also be used to formulate a linear ``UBOS'' model, in which the density field is directly related to image difference data and reconstructions are performed in a single step \cite{Grauer2020}.\footnote{Note that a non-linear variant of UBOS can be used to perform time- and scale-resolved reconstructions, but the linear technique is required to reconstruct mean (or otherwise filtered) fields from mean (or similarly filtered) data.}\par

In this work, we perform deflection sensing with a version of the ONERA FOLKI algorithm \cite{Champagnat2011} that has been modified for BOS \cite{Plyer2016, Nicolas2016}. FOLKI is an extension of Lukas--Kanade optical flow; in the classical Lukas--Kanade method, Eq.~\eqref{equ:linear optical flow} is converted into a well-posed problem by assuming the deflection field is locally-uniform within a user-defined interrogation window. FOLKI augments the Lukas--Kanade algorithm with a coarse-to-fine multi-resolution analysis using Gaussian pyramids for interpolation. The resulting non-linear functional is minimized via Gauss--Newton optimization. In addition to FOLKI, we use the UBOS formulation \cite{Grauer2020} to simultaneously solve the optical flow and tomographic reconstruction problems, as described in Sec.~\ref{sec:BOS:reconstruction}. UBOS eliminates the need for a non-physical closure at the deflection sensing stage, but it also requires a tailored experimental setup. Section~\ref{sec:modeling:forward} includes a discussion of deflection sensing and UBOS in the context of large deflections and depth-of-field effects.\par

\subsection{Refraction with a Finite Aperture}
\label{sec:BOS:continuous models}
\begin{figure*}[t]
    \centering
    \includegraphics[width=5in]{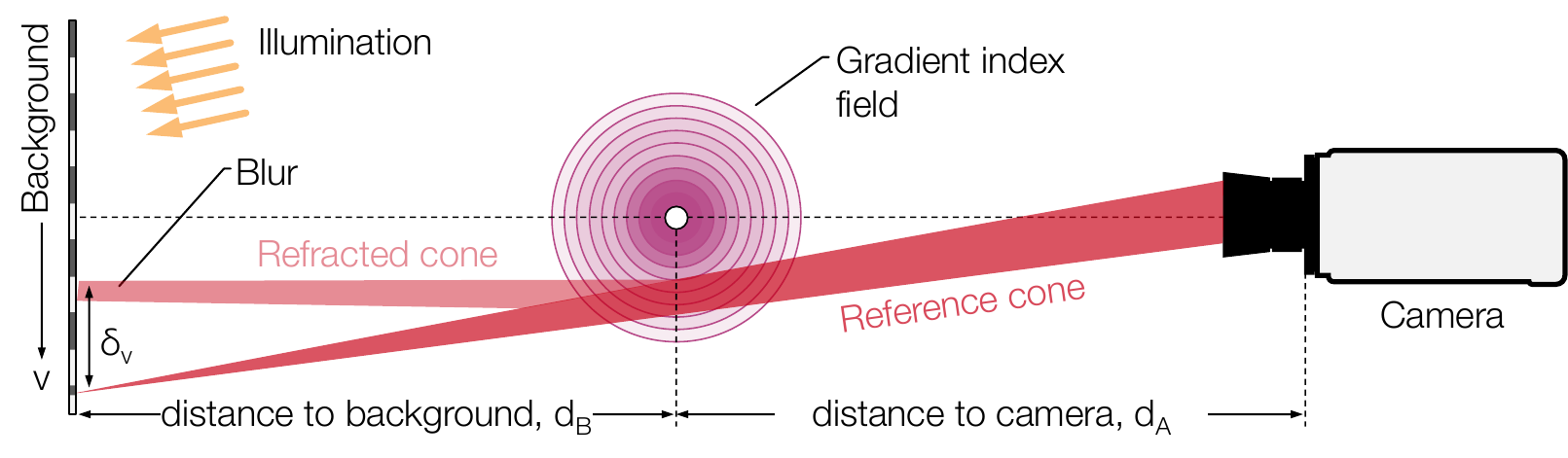}
    \caption{Illustration of a BOS setup that indicates the frustum accepted by the aperture; diagram is not to scale.}
    \label{fig:BOS}
\end{figure*}

Visible distortions of the background pattern arise when wavefronts of light bend across refractive index gradients. The speed of light in a gas of uniform composition is a linear function of the local molecular density, as codified by the Gladstone--Dale equation \cite{Gardiner1981}. In the limit of geometric optics, the propagation of light can be approximated by infinitesimal ``rays'' that travel normal to phase fronts of the electromagnetic field, and refraction may be modeled in terms of a path integral along these rays \cite{Stam1996, Settles2001}. Figure~\ref{fig:BOS} depicts the propagation of light from the background pattern to the camera, where the reference and refracted cones comprise rays that emanate from a focal point on the background plate. Once the flow has been introduced, rays from a range of points, corresponding to a contiguous region on the background screen, may refract into the collection frustum, distorting and also blurring the pattern.\par

The picture in Fig.~\ref{fig:BOS} is commonly simplified by considering a pinhole camera, which has a point (0D) aperture such that the cones in Fig.~\ref{fig:BOS} collapse to ``thin'' (1D) rays. Taking this approach, the $\alpha$-direction deflection of the reference ray, for $\alpha \in \{u, v\}$, is given by
\begin{equation}
    \delta_\alpha \approx \underbrace{M\frac{d_\mathrm{B}\,\psi\,G}{n_0}}_{C_\mathrm{sys}} \int_\mathrm{ray} \frac{\partial \rho}{\partial \alpha} \,\mathrm{d}s,
    \label{equ:ray equation:thin}
\end{equation}
as derived in \cite{Settles2001, Raffel2015, Grauer2020} and elsewhere. Here, $s$ is a progress variable that denotes the distance along a ray, and the system constant, $C_\mathrm{sys}$, includes the distance from the flow to the background, $d_\mathrm{B}$, Gladstone--Dale constant, $G$, pixel pitch, $\psi$, ambient refractive index, $n_0$, and lens magnification,
\begin{equation}
    M = \frac{f}{d_\mathrm{B} + d_\mathrm{A} - f},
\end{equation}
where $d_\mathrm{A}$ is the distance from the camera to the flow. Equation~\eqref{equ:ray equation:thin} assumes that the image plane is coincident with the background plate, where the $x$  and $y$ axes indicate horizontal and vertical deflections, and derivatives of the density field are computed in this coordinate system.\footnote{It should be noted that $\partial/\partial\alpha$ is a physical-space partial derivative that is aligned with the $\alpha$-direction of the image plane. See \cite{Grauer2018} for the general 3D case with an arbitrary angle between the image plane and background plate.} Furthermore, while refraction causes rays to bend, the refracted trajectory in Eq.~\eqref{equ:ray equation:thin} remains nearly straight within the flow domain \cite{Goldhahn2007, Atcheson2012}. Hence, density gradients may be integrated along the straight reference path when calculating $\delta_\alpha$ instead of following the true, ever-so-slightly curved path. This simplification, which spares the cost of non-linear ray tracing, is called the paraxial approximation.\par

Unfortunately, errors associated with the pinhole model can be significant \cite{Walsh2000, Kirmse2011, Amjad2020}. Real cameras have a finite aperture that accepts a cone of light that may traverse a range of refractive index gradients, as illustrated by the concentric rings in Fig.~\ref{fig:BOS}. In the flow-off condition, light originating at the background plate is assumed to be in focus. Differential refraction through the flow manifests as blur in the flow-on image. The base of the cone corresponds to the camera aperture, of diameter $D_\mathrm{ap}$, which can be calculated from the lens' $f$-number: $f/D_\mathrm{ap}$, where $f$ is the focal length. The approximation in Eq.~\eqref{equ:ray equation:thin} is relaxed by integrating over the cone,
\begin{equation}
    \delta_\alpha \approx \frac{4 C_\mathrm{sys}}{\pi D_\mathrm{ap}^2} \int_0^{D_\mathrm{ap}/2}\int_0^{2\pi} \left(  \int_{\mathrm{ray}(l,\theta)} \frac{\partial \rho}{\partial \alpha} \,\mathrm{d}s \right) r \,\mathrm{d}\theta \,\mathrm{d}l,
    \label{equ:ray equation:cone}
\end{equation}
where $l$ and $\theta$ are polar coordinates that span the aperture, itself centered at $l = 0$~mm in this formulation. Each ray in the inner integral stretches from a point on the aperture, $(l,\theta)$, to the focal point of a pixel, which can be calculated by inverting the camera transform and tracing the pixel's chief ray to the camera's focal plane \cite{Grauer2023}. Equation~\eqref{equ:ray equation:cone} is the implicit model that underpins the aperture sampling technique of Cook and Porter \cite{Cook1984}, which was originally developed to simulate depth-of-field effects in computer graphics applications. It should also be noted that the \textit{forward} Monte Carlo simulations of BOS imaging in Rajendran et al. \cite{Rajendran2019a} are consistent with Eq.~\eqref{equ:ray equation:cone} since all the light effectively originates at the background plate.\par

Equation~\eqref{equ:ray equation:cone} accounts for the finite depth of field associated with a 2D aperture. Hence, we expect this expression to be \textit{truer to life} than the pinhole model in Eq.~\eqref{equ:ray equation:thin}, which assumes a 0D aperture. Consequently, inverting BOS data with Eq.~\eqref{equ:ray equation:cone} instead of Eq.~\eqref{equ:ray equation:thin} should enhance the accuracy of reconstructed density fields as well as analysis thereof. However, from our reading of the literature, \textit{all existing algorithms for BOS tomography utilize a pinhole model}.\par

\subsection{Neural-Implicit Reconstruction Technique}
\label{sec:BOS:reconstruction}
\begin{figure*}[ht]
    \centering
    \includegraphics[width=5in]{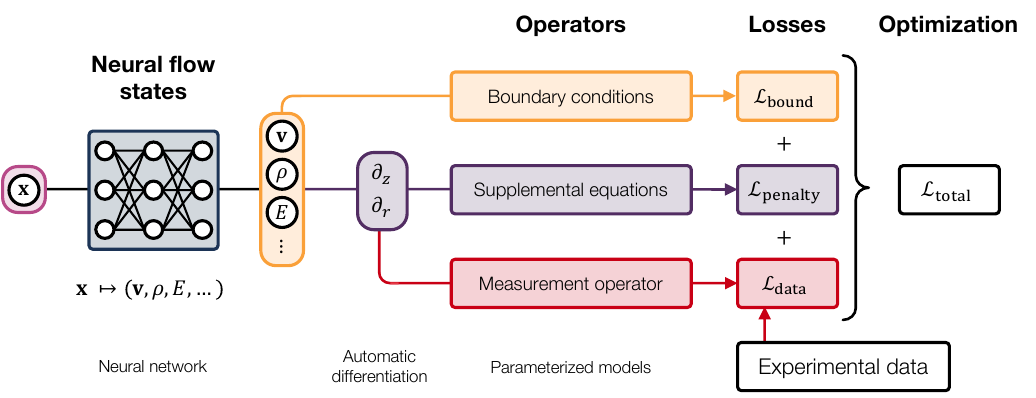}
    \caption{NIRT allows differential equations and boundary conditions to be included in the reconstruction.}
    \label{fig:arch}
\end{figure*}

Given a set of deflection data and a suitable forward model, a tomography algorithm may be used to invert said model and thereby recover the unknown density field corresponding to said data. This is an ill-posed inverse problem, meaning that additional information is needed to obtain a unique, stable, and physically-plausible solution from a real measurement \cite{Daun2016}. Many techniques have been developed to regularize the inversion, including iterative algorithms, classical regularization methods, and comprehensive data assimilation schemes \cite{Grauer2023}. We use a neural-implicit reconstruction technique (NIRT), depicted schematically in Fig.~\ref{fig:arch}, where the density field is represented by a continuous function in the form of a deep, feed-forward neural network,
\begin{equation}
    \mathcal{N}\mathopen{}\left(\boldsymbol\uptheta\right): \mathbf{x} \mapsto \rho.
    \label{equ:map}
\end{equation}
In this paper, $\mathbf{x} = [r,z]^\top$ is a vector of cylindrical coordinates and $\boldsymbol\uptheta$ contains all the free parameters in $\mathcal{N}$, which determine the mapping from $\mathbf{x}$ to $\rho$. We implement a hard positivity constraint on $\rho$ by parameterizing it with an exponential function. Further details about $\mathcal{N}$ can be found in \ref{app:network}. Per \ref{app:data assimilation}, the outputs of $\mathcal{N}$ can be expanded to include additional variables, like velocity and total energy, which facilitates multi-modal, multi-physics data assimilation. Apart from \ref{app:data assimilation}, we follow the procedure and parameterization described in this section.\par

Reconstruction amounts to optimizing $\boldsymbol\uptheta$ to minimize an objective loss that includes measurement, penalty, and boundary terms,
\begin{equation}
    \hat{\boldsymbol\uptheta} = \arg\underset{\boldsymbol\uptheta}{\min} \;\mathcal{L}_\mathrm{meas} + \lambda_1 \mathcal{L}_\mathrm{penalty} + \lambda_2 \mathcal{L}_\mathrm{bound}.
    \label{equ:optimization}
\end{equation}
The first of these, $\mathcal{L}_\mathrm{meas}$, compares experimental data to synthetic measurements computed using the density field from $\mathcal{N}$, e.g., via Eq.~\eqref{equ:ray equation:thin} or \eqref{equ:ray equation:cone}. A penalty may also be specified to discourage undesirable characteristics in $\rho$, such as high-spatial-frequency content, and a boundary loss can be added to enforce known inflow, outflow, symmetry, or wall conditions. Similar to many optical flow techniques, the influence of $\mathcal{L}_\mathrm{penalty}$ and $\mathcal{L}_\mathrm{bound}$ are controlled by $\lambda_1$ and $\lambda_2$, which must be carefully selected to ensure a proper balance of the data with the additional assumptions about the density field encoded in $\mathcal{L}_\mathrm{penalty}$ and $\mathcal{L}_\mathrm{bound}$. These parameters may be optimized through conventional approaches, such as the widely used L-curve analysis of Hansen and O'Leary \cite{Hansen1992}, employed in this work, or more modern techniques introduced in the machine learning community, such as gradien-t \cite{Wang2021} or neural-tangent-kernel \cite{Wang2022} adaptive weighting schemes. The network is implemented in a differentiable computing environment like TensorFlow or PyTorch, and Eq.~\eqref{equ:optimization} is executed with a backpropagation algorithm. NIRT algorithms akin to this one have been demonstrated for X-ray \cite{Sun2021, Zhang2021, Zang2021, Ruckert2022}, chemiluminescence \cite{Zhang2023}, generic emission \cite{Kelly2023, Kelly2024}, and BOS \cite{Molnar2023a, Molnar2023b} tomography.\par

As mentioned in Sec.~\ref{sec:BOS:optical flow}, we explore two formulations of the measurement loss. First, we reconstruct the density field from the deflection data we obtain using FOLKI. Second, we directly invert the image data via UBOS. In the first instance, the data loss is
\begin{equation}
        \mathcal{L}_\mathrm{meas} = \frac{1}{\sqrt{2} \left|\mathcal{I}\right|} \iint_\mathcal{I} \left\lVert \boldsymbol\updelta_\alpha^\mathrm{OF} - \boldsymbol\updelta_\alpha^\mathrm{NN} \right\rVert^2 \mathrm{d}\mathcal{I},
        \label{equ:OF data loss}
    \end{equation}
where $\mathcal{I}$ is the image domain, $\lVert\cdot\rVert$ is the Euclidean norm, and the superscripts OF and NN indicate deflections from the optical flow algorithm and neural network, respectively. Integrals over $\mathcal{I}$ are approximated by summing the integrand at each pixel (or a random subset of the pixels). The UBOS data loss is
\begin{equation}
        \mathcal{L}_\mathrm{meas} = \frac{1}{\left| \mathcal{I} \right|} \iint_\mathcal{I} \left(\boldsymbol\updelta^\mathrm{NN} \cdot \nabla I + \frac{\partial I}{\partial t}\right)^2 \mathrm{d}\mathcal{I},
        \label{equ:UBOS data loss}
    \end{equation}
where $\nabla$ is implemented with a second-order central differencing scheme and $\partial I/\partial t \approx I(t + \Delta t) - I(t)$. In both conventional and unified BOS, the elements of $\boldsymbol\updelta^\mathrm{NN}$ are computed using Eq.~\eqref{equ:ray equation:thin} or \eqref{equ:ray equation:cone} for thin- or cone-ray reconstructions, respectively. Section~\ref{sec:modeling:inverse} demonstrates the superiority of the cone-ray approach, which is predicated upon a more accurate model of the imaging system.\par

In addition to the measurement loss, we employ a total variation (TV) penalty,
\begin{equation}
    \mathcal{L}_\mathrm{penalty} = \frac{1}{\left|\mathcal{V}\right|} \iiint_\mathcal{V} \left\lVert \nabla \rho \right\rVert \mathrm{d}\mathcal{V},
    \label{equ:penalty}
\end{equation}
where $\mathcal{V}$ is the measurement volume and $\nabla$ is the gradient operator in world coordinates. The TV loss promotes piecewise smoothness, which is desirable for high-speed flows due to the presence of sharp interfaces like shocks. If $\mathcal{L}_\mathrm{penalty}$ is omitted (or $\lambda_1$ is vanishingly small), the resultant density field will be subject to implicit regularization by the network architecture and optimization algorithm. This can lead to undesirable effects like over-smoothing or high-frequency artifacts associated with a Fourier encoding. Therefore, it is preferable to use a large network with sufficient ``expressivity'' to represent the true density field and, at the same time, to introduce an explicit source of regularization with known characteristics, such as TV or physics-based penalties in Eq.~\eqref{equ:penalty}. Throughout this work, we specify an explicit regularization term. A boundary loss is also included to account for the inflow conditions,
\begin{equation}
    \mathcal{L}_\mathrm{bound} = \frac{1}{\left|\mathcal{A}\right|} \iint_\mathcal{A} \left\lVert \rho - \rho_0 \right\rVert^2 \,\mathrm{d}\mathcal{A},
    \label{equ:boundary}
\end{equation}
where $\mathcal{A}$ denotes the inlet boundary and $\rho_0$ is an estimate of the free-stream density. Lastly, we note that the physical-space integrals in Eqs.~\eqref{equ:ray equation:thin}, \eqref{equ:ray equation:cone}, \eqref{equ:penalty}, and \eqref{equ:boundary} are approximated by Monte Carlo sampling.\par

\section{Flow Cases} 
\label{sec:cases}
Two scenarios are considered in this work. First is a numerical evaluation of BOS that features buoyancy-driven turbulent flow. This case is used to comprehensively assess the forward model. Second is a numerical and experimental study of hypersonic flow over a spherical bluff body, which has been well characterized in the literature. Both forward and inverse methods are tested in the second scenario to provide empirical support for our model.\par

\subsection{Buoyancy-Driven Turbulence}
\label{sec:cases:buoyancy}
\begin{figure*}[ht]
    \centering
    \includegraphics[width=5.5in]{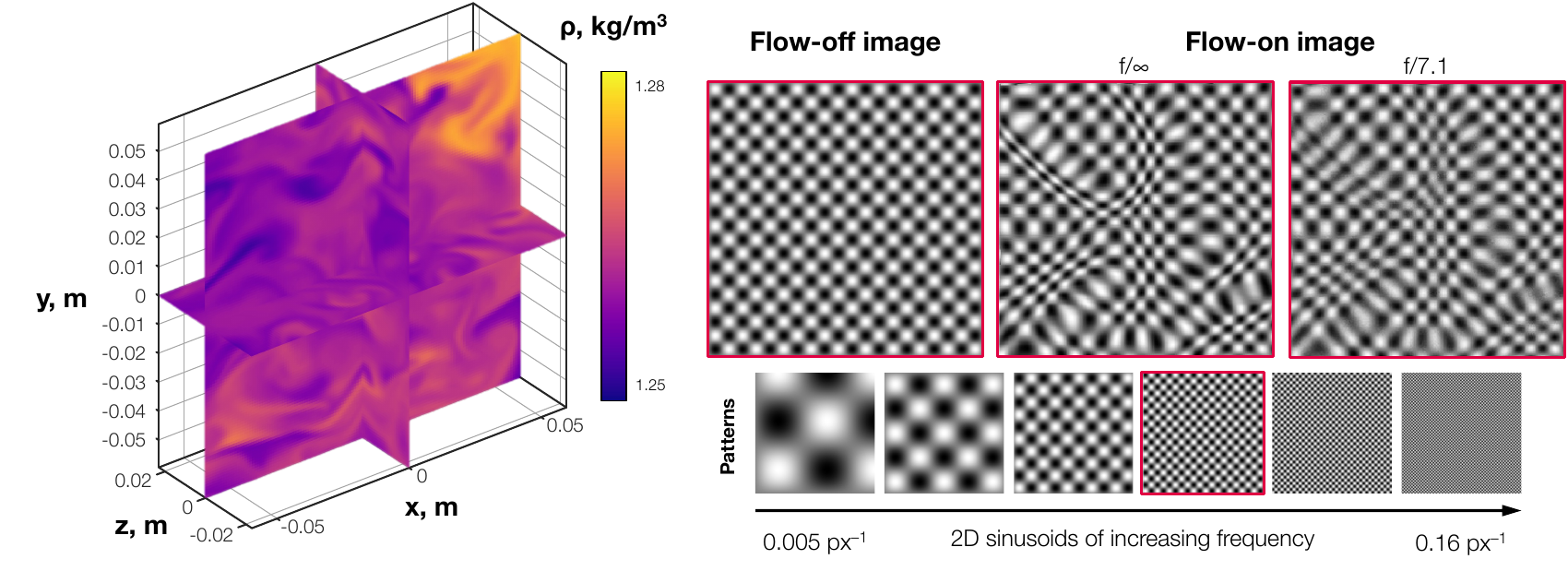}
    \caption{Buoyancy-driven flow: (left) snapshot of density from a DNS and (right) synthetic BOS images of the flow based on a sine-wave pattern.}
    \label{fig:DNS}
\end{figure*}

Density gradients in buoyancy-driven turbulence can exhibit a wide range of spatial scales, which is ideal for testing BOS. We simulate BOS measurements of buoyancy-driven turbulence using the direct numerical simulation (DNS) of Livescu et al. \cite{Livescu2014}, which is available in the Johns Hopkins Turbulence Database \cite{Li2008}. The simulation was conducted on a $1024^3$ periodic grid, with Schmidt and Froude numbers of unity and an Atwood number of 0.05. The miscible, two-fluid, incompressible Navier--Stokes equations were solved by the Los Alamos variable density code. The mixture was initialized at a random state in which both fluids were at rest. In this flow, buoyancy forces drive mixing, which quickly becomes turbulent and persists until the fluids are homogenized, after which the turbulence dies out. We select a subsection of this simulation, beginning at the onset of turbulent mixing and carrying on to its decay. A characteristic snapshot of the density field from this data set is shown on the left side of Fig.~\ref{fig:DNS}. Note that, although this is an incompressible flow, the composition gradients give rise to refractive index gradients that ultimately lead to deflections.\par

We dimensionalize the DNS data to simulate physically-plausible BOS measurements. The light fluid is assumed to be air, the heavy fluid is oxygen, and the temperature is 20~$^{\circ}$C throughout the domain. Although this mixture corresponds to Schmidt and Atwood numbers of 0.88 and 0.0435, which are slightly smaller than the respective DNS values of 1 and 0.05, these differences are within reason. We use the mean kinematic viscosity of an air--O$_2$ mixture, Froude number of 1, and turbulent Reynolds number of 17,765 to determine appropriate velocity, length, and time scales for the flow. A $12\times12\times4.7$~cm$^3$ subsection of the domain is selected for imaging; the Atwood number and local dimensionless density are used to determine the mass fraction of air and O$_2$, $Y_\mathrm{air}$ and $Y_\mathrm{O_2}$, and calculate the refractive index, $n = 1 + \rho \,(Y_\mathrm{air} G_\mathrm{air} + Y_\mathrm{O_2} G_\mathrm{O_2})$, where $\rho$ is the dimensional density, $G_\mathrm{air} = 2.26 \times 10^{-4}$~m$^3$~kg$^{-1}$, and $G_\mathrm{O_2} = 1.89 \times 10^{-3}$~m$^3$~kg$^{-1}$. Note that we amplify $G_\mathrm{O_2}$ by a factor of ten to generate the desired range of deflections -- up to 26~px -- for our narrow-aspect-ratio (quasi-2D) domain. Such large deflections allow us to examine BOS operability across a wide range of conditions in Sect.~\ref{sec:modeling:forward}. Gradients of the refractive index field are computed by second-order finite differencing.\par

Synthetic measurements are generated for a virtual camera with a $1024 \times 1024$~px sensor and 3.45~$\upmu$m pixel pitch. The camera is placed 1~m from the center of the flow and 2~m from the background plate. We specify a 35~mm lens, resulting in a field of view close to $10\times10$~cm$^2$ in the flow domain. Ray tracing is conducted using a fourth-order Runge--Kutta scheme; 100 $(l, \theta)$ samples are drawn for each pixel to approximate Eq.~\eqref{equ:ray equation:cone}, as discussed in \cite{Molnar2023a, Molnar2023b}. Images are generated at a frame rate of 40~Hz for three seconds using aperture settings of $f/\infty$ and $f/7.1$. The finite aperture setting, circle of confusion, and peak deflections are configured to match those of a prominent BOS experiment \cite{Nicolas2017a}. Reference and deflected images are generated using a series of background patterns, including a pseudo-random dot pattern and 2D sinusoids of increasing frequency. The sine-wave pattern is employed to assess the linear formulation of optical flow. We test background frequencies of 4.9~$\times \ 10^{-3}$, 9.8~$\times \ 10^{-3}$, 19.5~$\times \ 10^{-3}$, 39.1~$\times \ 10^{-3}$, 78.1~$\times \ 10^{-3}$, and 156.3~$\times \ 10^{-3}$~px$^{-1}$ in this work; these patterns are displayed in the bottom-right quadrant of Fig.~\ref{fig:DNS}. Sample images based on the same flow instance are shown in the top-right quadrant. These images exhibit noticeable distortions caused by refraction through the flow. Depth-of-field-induced blur is also apparent in the $f/7.1$ image, which can be seen in the highlighted region.\par

\subsection{Hypersonic Flow Over a Sphere}
\label{sec:cases:sphere}
\subsubsection{Experimental Facility, Setup, and Data Sets}
\begin{figure*}[ht]
    \centering
    \includegraphics[width=6.5in]{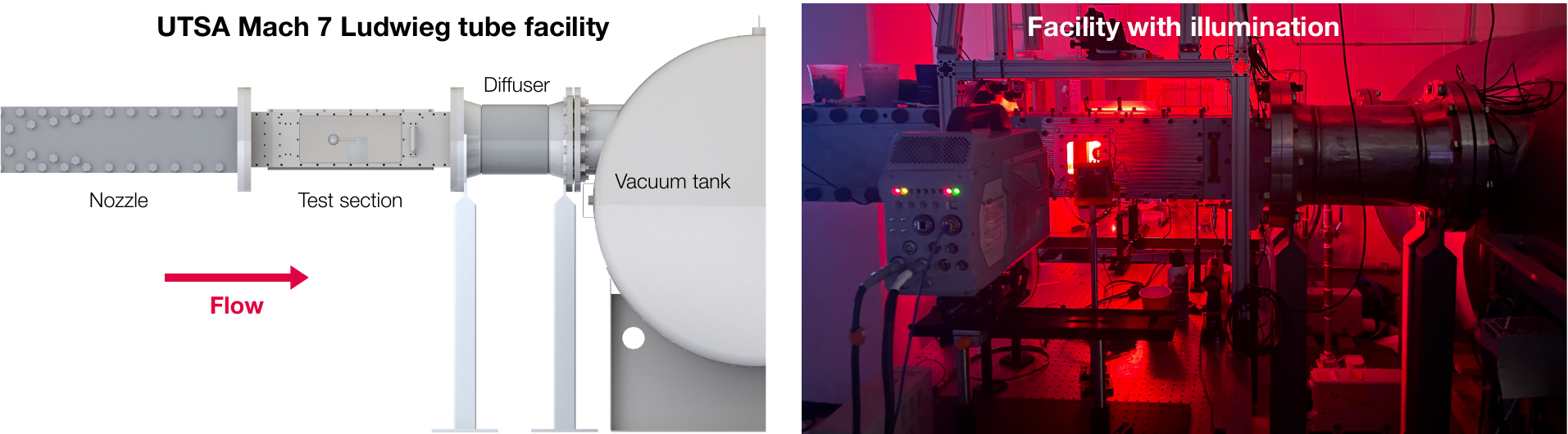}
    \caption{UTSA Mach~7 Ludwieg tube: (left) model of the facility and (right) photo of the experimental setup.}
    \label{fig:facility}
\end{figure*}

A well-characterized canonical flow with sharp density gradients is needed to verify the performance of our cone-ray model. Therefore, we test our method using BOS measurements of hypersonic flow over a spherical bluff body. All testing for this work was conducted at The University of Texas at San Antonio (UTSA) Mach~7 wind tunnel, pictured in Fig.~\ref{fig:facility}. This facility comprises a Ludwieg tube that can operate at stagnation conditions up to 13.8~MPa and 700~K at Mach~$7.2 \pm 0.2$. A wide range of viable operating conditions enables Reynolds numbers up to $200 \times 10^6$~m$^{-1}$ to be achieved across multiple steady-state passes, each lasting between 70--100~ms. The $203 \times 203$~mm$^2$ test section can be configured to provide optical access via acrylic or fused-silica windows. Detailed characterization of the facility operating conditions can be found in \cite{Hoffman2023}, and sub-component analyses are reported in the works of Bashor, Hoffman, and their coworkers \cite{Bashor2019, Hoffman2020, Hoffman2021}.\par

We perform BOS measurements of flow over a steel sphere of diameter $D = 50.8~\mathrm{mm}$ ($Re_\mathrm{D} \approx 1.1 \times 10^6$). The sphere is fastened to an optical post that we strut mount to the tunnel floor. Optical access is provided by two fused-silica windows with a $101.6\times101.6$~mm$^2$ cross section and thickness of $12.7$~mm. The right side of Fig.~\ref{fig:facility} shows the camera, which is aimed at the measurement volume and positioned 0.5~m from the centerline of the tunnel (as measured from the lens). A back-lit background plate is fixed to the rear tunnel window, i.e., 0.2~m from the centerline and 0.7~m from the camera, by a custom additively-manufactured mount. Since the plate is attached to the tunnel, oscillations of the test section may be deduced from the flow-on image set; Hoffman et al. \cite{Hoffman2023} report minimal vibrations, although tunnel recoil does occur. We print a 2D sine-wave pattern with a period of 1.4~mm on a sheet of transparency that we insert in the mount. This setup allows for rapid fabrication and testing of alternative patterns. Back illumination is provided by a high-powered LED which shines through an aspheric lens, Fresnel lens, and frosted acrylic sheet, generating diffuse, near-uniform illumination of the pattern. This assembly is positioned 0.165~m from the plate. Full power is provided to the LED to avoid flickering, and a neutral density filter is inserted between the diode and aspheric lens to avoid saturation of the camera sensor.\par

Imaging is conducted with a high-speed camera (Photron, FASTCAM SA-Z) at a rate of 20~kHz with an exposure time of 10~$\upmu$s. The camera has a monochromatic 1~MP sensor and pixel pitch of 20~$\upmu$m. We outfit the camera with a 100~mm macro lens (ZEISS, Milvus 2/100M). For each test case, 3000 images are recorded before the diaphragm bursts; these images are used to construct a high-fidelity reference signal. Next, 1400 flow-on frames from the first steady-state tunnel pass are averaged to obtain the mean deflected image. In order to assess our cone-ray model, we perform tests with six aperture settings, ranging from $f/4$ to $f/22$, with corresponding circles of confusion at the centerline of 6.25--1.14~mm. We manually align the focal plane and background plate using four sharp, black-bordered dots, printed at the corners of our pattern. Illumination is provided by a high-power red LED (Luminus Devices, CBT-90-RX-L15-BN101) that is triggered by the camera.\footnote{An epiphenomenal effect of our system of illumination is an iridescent bath of burgundy--indigo light that fills the laboratory, as captured in the photo on the right side of Fig.~\ref{fig:facility}.} The LED is activated by a falling-edge signal from the camera that triggers a 100~ns top-hat pulse in the middle of the camera's exposure. This period sustains a maximum fluid-element displacement of less than 1~px, which is sufficient to freeze the flow. Further details regarding the system setup, calibration, and data acquisition protocols are provided in \cite{Molnar2024}.\par

\begin{figure*}[ht]
    \centering
    \includegraphics[width=6.5in]{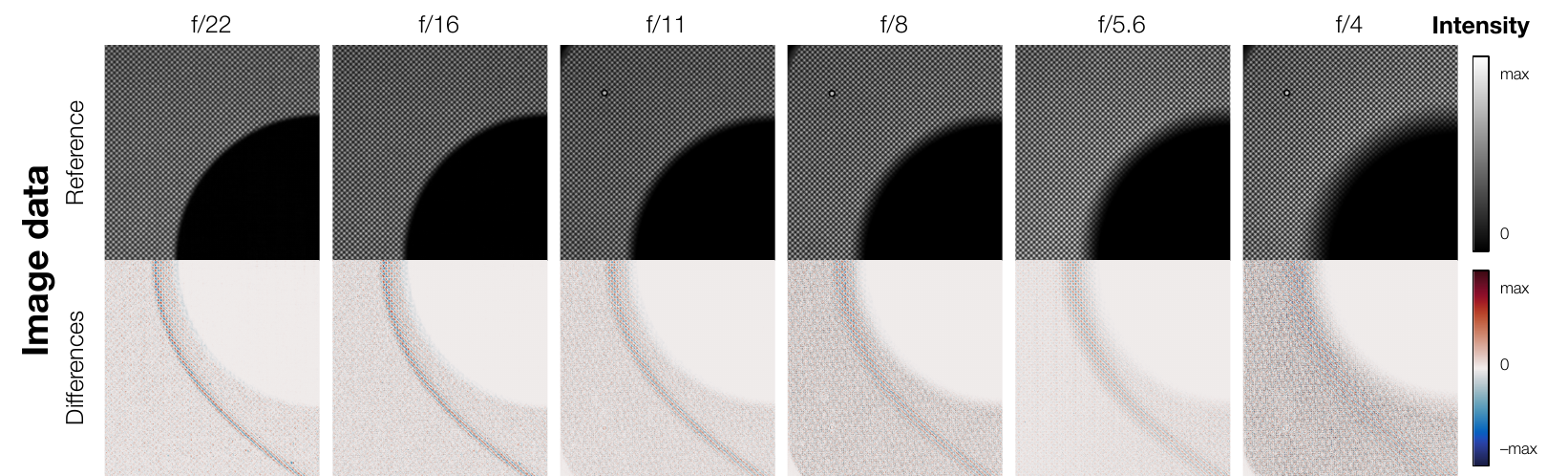}
    \caption{Rows depict mean flow-off images and image difference data for each test; columns indicate the $f$-number.}
    \label{fig:image data}
\end{figure*}

Figure~\ref{fig:image data} depicts a representative set of images from the sphere flow experiment. The top row contains reference, a.k.a. flow-off, images that were recorded with $f$-numbers ranging from $f/22$ to $f/4$. In the context of BOS, these aperture settings span the very small to the very large. Scanning the figure from left to right, as the aperture is opened, the silhouette of the spherical test article morphs from a sharp relief to a blurry smudge, while the background remains in focus across the board. Corresponding ``image differences'' -- flow-on minus flow-off data -- are shown in the bottom row of Fig.~\ref{fig:image data}. All six images are recorded at comparable test conditions, so discrepancies between the columns must be caused \textit{solely} by the camera's decreasing depth of field (i.e., with decreasing $f$-number). In other words, although a sharp bow shock is visible in the $f/22$ column, the same shock in the $f/4$ image difference appears to be hazy and broadened. Nevertheless, given the right imaging model, it should be possible to better estimate the underlying density field from any of these data sets.\par

\subsubsection{Numerical Data Set}
\label{sec:cases:sphere:CFD}
An inviscid, compressible CFD simulation of the sphere flow case is conducted to compare the output of our forward model to our experimental data sets. We perform the simulation in SU2~7.3.0 \cite{Economon2016} using a domain size of $1.5D$, axially, and $2D$, radially, to prevent shock reflections. The grid consists of around 400,000 unstructured triangular cells that are body-fitted to the sphere. This cell count was found to be more than sufficient for grid convergence in our previous work on a similar flow \cite{Molnar2023a}. The advection upstream splitting method is used for convective fluxes, and gradients are computed via inverse distance weighted least squares. Experiments and simulations of the tunnel have shown approximately-uniform flow conditions in the region surrounding our test article \cite{Hoffman2023}. Therefore, we set the inflow conditions of our simulation to those reported in \cite{Hoffman2023}. Since the bow shock is the primary flow feature of interest in this work, we simulate axisymmetric flow over a hemisphere and neglect the wake region.\par

\section{Depth-of-Field Effects in BOS}
\label{sec:modeling}
Although BOS should be conducted with the largest possible $f$-number, in high-speed tests, it is sometimes necessary to open up the aperture to allow more light into the camera. This reduces the minimum exposure time needed to achieve a prescribed SNR, thereby helping to freeze the flow. However, opening the aperture also reduces the camera's depth of field, causing blurry images like those seen at the right side of Fig.~\ref{fig:image data} and in the BOS literature \cite{Kirmse2011, Elsinga2004}. Such data are poorly-modeled by a pinhole camera. In this section, we analyze the implications of infinitesimal ``thin'' rays and realistic ``cone'' rays in the forward and inverse modeling of BOS.\par

\subsection{Forward Modeling}
\label{sec:modeling:forward}
We first assess linear optical flow for BOS using our simulated measurements of buoyancy-driven mixing. Linear optical flow corresponds to the first-order Taylor expansion in Eq.~\eqref{equ:Taylor series}, which is valid for low-spatial-frequency background patterns. This equation is perfectly satisfied in the limiting case of a background that exhibits a constant linear slope in intensity, e.g., the ``patterns'' tested by Mier and Hargather \cite{Mier2016}. Unfortunately, the spatial frequency of a BOS pattern is directly tied to SNR because image difference data, like the images in the bottom row in Fig.~\ref{fig:image data}, is proportional to gradients of the background pattern, $\nabla I$. Conversely, while large intensity gradients produce visible deflections, the linear approximation is quickly violated when the magnitude of $\nabla I$ falls below that of $\boldsymbol\updelta$, leading to a more challenging, non-linear optical flow problem that is based on the implicit mapping in Eq.~\eqref{equ:intensity constancy}. Moreover, many BOS experiments require time-averaging to obtain a reliable signal, but the aforementioned non-linearities interact with blur and turbulent fluctuations. This interaction means that individual flow-on images, which have a much lower SNR than the mean flow-on image, must be processed one-by-one, producing a set of instantaneous deflections that may be averaged to obtain the mean deflection field. Such data sets are often replete with biased errors, as a result, especially close to shocks and in regions of the image associated with intense turbulence. In other words, although it is more reliable to process a single pair of \textit{time-averaged} flow-on and flow-off images than to average a set of noisy deflection snapshots, the former approach is only valid in the \textit{linear regime of optical flow}.\par

\begin{figure*}[t]
    \centering
    \includegraphics[width=4in]{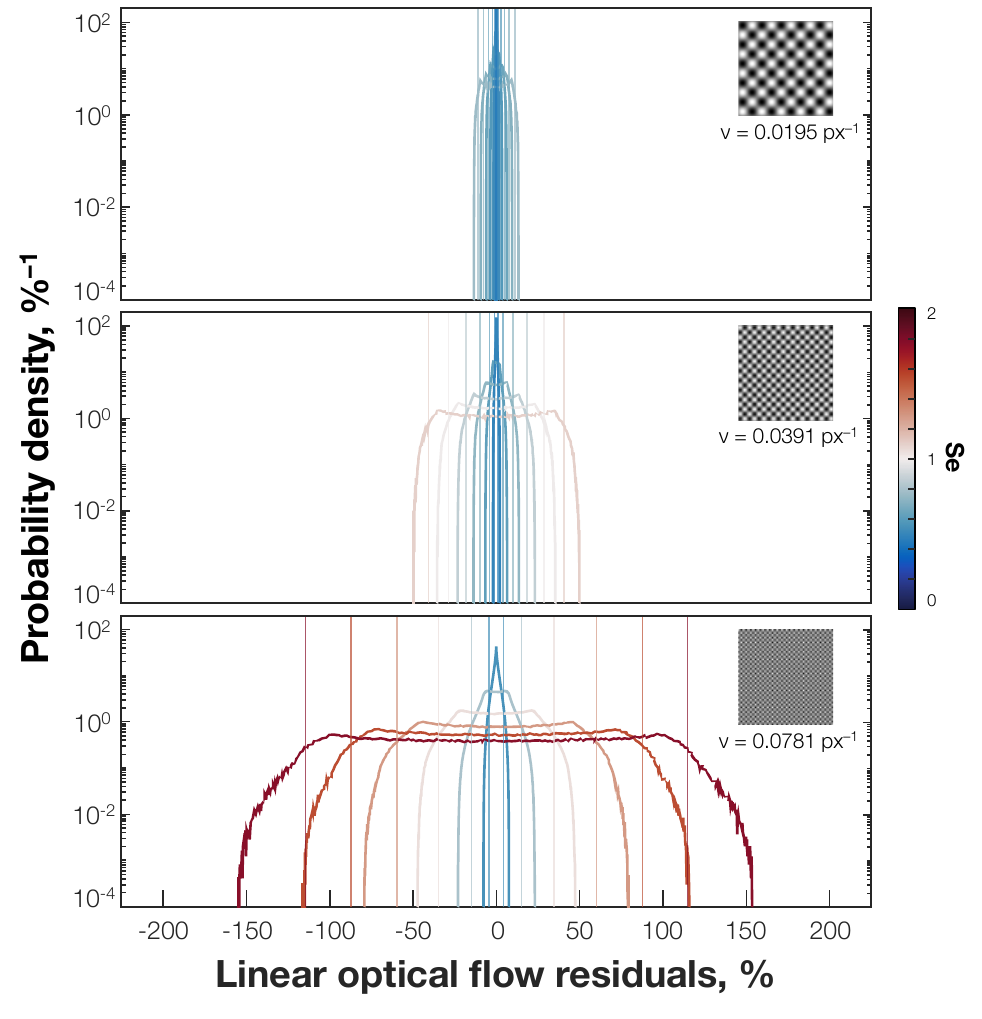}
    \caption{PDFs of residuals from Eq.~\eqref{equ:Taylor series} for three sine-wave background patterns.}
    \label{fig:OF errors:pinhole}
\end{figure*}

Synthetic BOS data are generated for the buoyancy-driven turbulent flow, using $f/\infty$ (pinhole) and $f/7.1$ configurations, to assess the validity of Eq.~\eqref{equ:Taylor series}. A reference image and set of time-resolved distorted images are produced for each of the sine-wave patterns in the bottom right of Fig.~\ref{fig:DNS}. Figure~\ref{fig:OF errors:pinhole} depicts probability density functions (PDFs) of the residuals from Eq.~\eqref{equ:Taylor series} for three of the pinhole tests. Results are shown for backgrounds of increasing spatial frequency, arranged in frequency from low (top) to high (bottom). Each line in Fig.~\ref{fig:OF errors:pinhole} corresponds to the distribution of residuals for deflections of a similar magnitude; the magnitude of deflections is indicated by color in terms of a dimensionless number that we call a Settles number,
\begin{equation}
    Se = 4 \nu \, \overline{\delta},
\end{equation}
where $\nu$ is a characteristic spatial frequency of the background pattern and $\overline{\delta}$ is the magnitude of deflections in reciprocal units (e.g., px$^{-1}$ and px, respectively). The factor of four corresponds to the length of the roughly-linear span of the sinusoid in between peaks and troughs. Note that $\nu$ depends on the physical pattern and optical setup of an experiment, and $\overline{\delta}$ depends on the setup as well as the flow. Vertical lines in Fig.~\ref{fig:OF errors:pinhole} mark out the 95\% confidence interval for the distribution of the same $Se$ (plotted in the same color). Since the distribution of deflections is constant across these tests, the only difference between the subplots in Fig.~\ref{fig:OF errors:pinhole} is the spatial frequency of the background pattern. Naturally, as this frequency (and hence $Se$) increases, the Taylor-expansion approximation breaks down and residuals in Eq.~\eqref{equ:Taylor series} grow large. An inflection point is apparent around $Se = 1$, after which there is a dramatic increase in the magnitude of residuals with the Settles number. Ergo, $Se \ll 1$ is a safe criterion for using a linear optical flow algorithm or the linear form of UBOS. The Settles number can be controlled through placement of the camera and background plate, which determine the magnitude of deflections per Eq.~\eqref{equ:ray equation:thin}, and by selecting a pattern with appropriate frequency content.\par

\begin{figure*}[ht]
    \centering
    \includegraphics[width=3.5in]{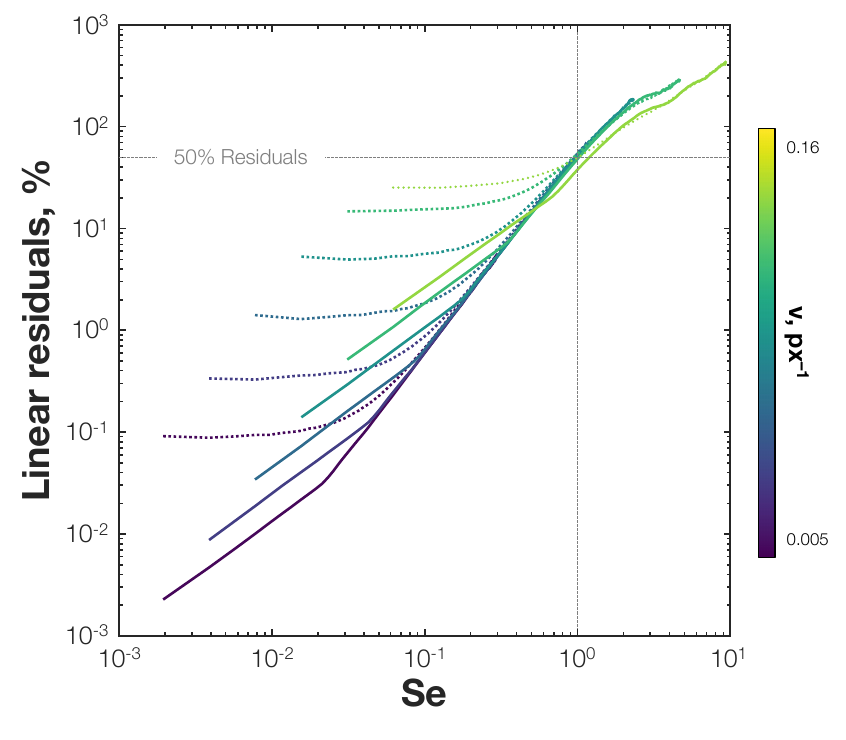}
    \caption{Width of 95\% confidence intervals for residuals from Eq.~\eqref{equ:Taylor series}, colored by the background-pattern frequency. Solid and dashed lines convey pinhole and cone-ray results.}
    \label{fig:OF errors:comparison}
\end{figure*}

Depth-of-field effects can also influence the validity of Eq.~\eqref{equ:Taylor series} because the underlying image formation process, given by Eq.~\eqref{equ:intensity constancy} is non-linear and does not commute with the area integral in Eq.~\eqref{equ:ray equation:cone}. Figure~\ref{fig:OF errors:comparison} depicts the width of the 95\% confidence interval for all six background patterns versus $Se$; solid lines indicate pinhole results and dashed lines are cone-ray results from the $f/7.1$ data set. To isolate the effects of blur, we condition the data set on the variance of deflections within a cone -- only pixels with a $\boldsymbol\updelta$-sample variance greater than 1~px$^2$ are included in the plot. Overall, the trend of 95\% confidence intervals versus $Se$ is consistent: there is a strong increase in residuals with increasing Settles number, and the lines collapse onto a single profile beyond $Se = 1$. However, at lower Settles numbers, depth-of-field effects limit the accuracy of Eq.~\eqref{equ:Taylor series}. This effect manifests as dotted lines (cone-ray residuals) ``tailing off'' with decreasing $Se$ compared to the pinhole results, which can satisfy Eq.~\eqref{equ:Taylor series} to arbitrary precision as $Se$ approaches zero. This result underscores our advice to keep the aperture small and frequency content of the pattern limited in order to facilitate accurate deflection sensing via linear optical flow.\par

Note that we have restricted our discussion in this section to the errors associated with simple linear optical flow schemes, and we do not address multi-resolution algorithms. In addition, while we consider sinusoidal patterns, one could define a Settles number for random-dot patterns, Gaussian noise, and other options, e.g., using the dominant frequency as determined by a 2D discrete Fourier transform.\par

\begin{figure*}[ht]
    \centering
    \includegraphics[width=6.5in]{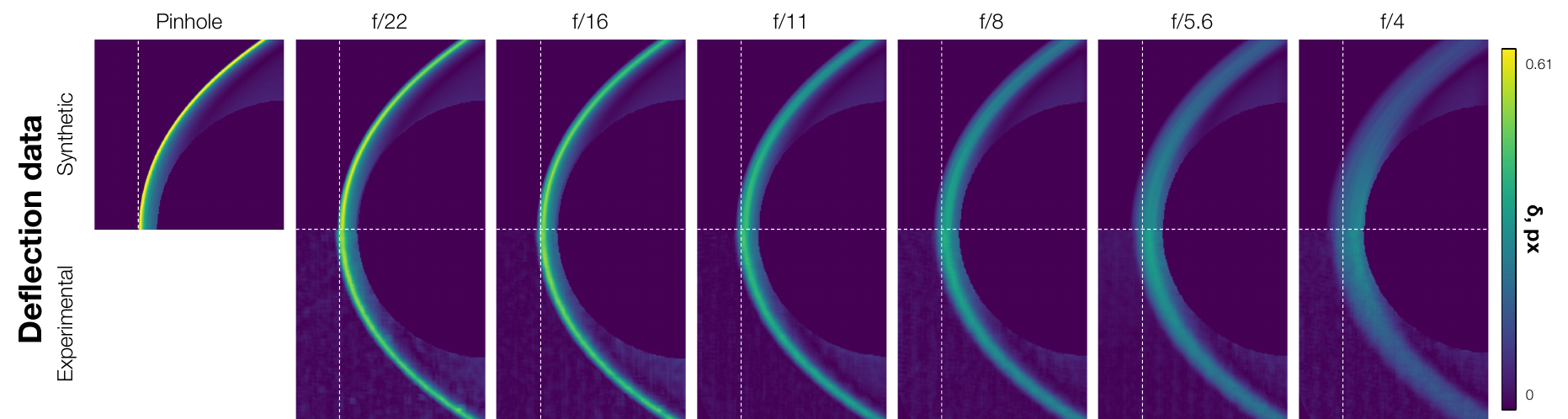}
    \caption{Synthetic (top) and experimental (bottom)  deflection magnitudes for hypersonic flow over a sphere; columns denote the aperture setting.}
    \label{fig:deflections}
\end{figure*}

To experimentally validate our forward BOS model, we compare deflections from the hypersonic sphere flow experiment, obtained with FOLKI, to our CFD predictions, computed using the simulation described in Sec.~\ref{sec:cases:sphere:CFD}. Experimental deflection data are shown in the bottom row of Fig.~\ref{fig:deflections}, and CFD-based deflections are shown in the top row. The trend with $f$-number is similar to the trend observed in Fig.~\ref{fig:image data}, namely: as the aperture is opened, the shock is smeared out and deflections become smaller in magnitude. Cone-ray deflections in the bottom row of Fig.~\ref{fig:deflections} correspond to Eq.~\eqref{equ:ray equation:cone}. To approximate this expression, we draw 1000 $(l, \theta)$ samples per pixel and conduct non-linear ray tracing along that line of sight via forward Euler integration.\footnote{Although Euler integration is less accurate than the fourth-order Runge--Kutta method, Euler integration with a small step size has been shown to strike the best balance between computational cost and accuracy \cite{Zigunov2023}.} These deflections are then averaged to get an effective value of $\boldsymbol\updelta$ for the cone. There is a strong qualitative correspondence between the experimental results and simulated deflections in Fig.~\ref{fig:deflections}. Moreover, peak deflections are listed in Table~\ref{tab:deflections} for quantitative comparison. With a few exceptions, possibly due to the effects of optical occlusions (discussed below), spurious deflections, and limitations of the CFD simulation, our experimental and synthetic deflections are in good agreement.\par

\begin{table*}[ht]
    \caption{Peak Deflections from Experiments and Simulations}
    \centering
    \tabcolsep=0.2cm
    \begin{tabular}{c c r r r r r r r}
     \hline\hline
     \\[-2ex]
     \multirow{2}{*}{Quantity} & \multirow{2}{*}{Source} & \multicolumn{7}{c}{Aperture Setting} \\ 
     &  &  \multicolumn{1}{c}{$f$/$\infty$}  & \multicolumn{1}{c}{$f$/22} & \multicolumn{1}{c}{$f$/16} & \multicolumn{1}{c}{$f$/11} & \multicolumn{1}{c}{$f$/8} & \multicolumn{1}{c}{$f$/5.6} & \multicolumn{1}{c}{$f$/4}\\
     \hline \\[-1.5ex]
     \multirow{ 2}{*}{$\delta_\mathrm{u}$} & Expt. &  & $0.544$    & $0.514$ &  $0.454$ & $0.388$ & $0.356$ & $0.309$ \\
     & CFD & $1.212$ & $0.590$ & $0.518$ & $0.419$ & $0.360$ & $0.299$ & $0.289$ \\
     \multirow{ 2}{*}{$\delta_\mathrm{v}$} & Expt. & & $0.403$ & $0.406$ & $0.287$ & $0.246$ & $0.262$ & $0.195$ \\
     & CFD & $0.680$ & $0.415$ & $0.415$ & $0.253$ & $0.249$ & $0.149$ & $0.121$ \\
     \multirow{2}{*}{$\lVert \boldsymbol\updelta \rVert$} & Expt. &  & $0.550$ & $0.521$ & $0.454$ & $0.388$ & $0.362$ & $0.309$ \\
     & CFD & $1.283$ & $0.611$ & $0.536$ & $0.419$ & $0.360$ & $0.302$ & $0.291$ \\
     \\[-2ex]
     \hline\hline
    \end{tabular}
    \label{tab:deflections}
\end{table*}

Notably, as the aperture is opened, the peak deflection produced by the bow shock moves towards the surface of the sphere. This effect is caused by the bluff body partially blocking cones of light at the periphery of the sphere. That is, as the collection frustum increases in size with the camera's $f$-number, the solid object blocks a larger portion of rays which correspond to nearby pixels. Occluded rays do not contribute to deflections, so there is an asymmetry in which rays close to the bow shock undergo significant refraction whereas rays on the other side of the cone hit the bluff body and thus do not contribute to the signal. The net effect is to artificially enhance deflections in this region compared to pixels further away from the body, whose full cone reaches the background pattern. Our forward model accurately predicts this part of the imaging process, which lends credence to the idea that our model can be used to correct for blur and ocular occlusions during inverse analysis.\par

\subsection{Inverse Analysis}
\label{sec:modeling:inverse}
Here, we implement a NIRT for inversion of BOS data in TensorFlow~2.10.1. Our networks consist of five hidden layers with 250 nodes apiece. Weights are randomly initialized with a standard normal distribution, and the biases start off at zero. We do not apply a hard-constraint on the axis of symmetry, but doing so could potentially reduce discrepancies near the centerline \cite{Lu2021}. To stabilize the reconstruction and reduce its computational cost, we perform a discrete, two-point Abel inversion of our deflection data. This yields a crude estimate of $\rho$: \textit{crude} because Abel inversion is highly sensitive to noise, particularly so near the centerline due to the singularity along that axis \cite{Molnar2023a, Sipkens2021}. Here, we merely use Abel inversion as a starting point. We pre-train the network for each data set (i.e., $f$-number) on the corresponding Abel inversion. That is, we use a single loss term which compares the output of $\mathcal{N}$ to the Abel-based density field. Pre-training is conducted with the Adam optimizer at a learning rate of $10^{-3}$ for two epochs followed by a rate of $10^{-4}$ for another two epochs, at which point the network closely approximates the Abel result. Next, we switch to Eq.~\eqref{equ:UBOS data loss} and repeat the same training sequence. Batch sizes for $\mathcal{L}_\mathrm{meas}$, $\mathcal{L}_\mathrm{penalty}$, and $\mathcal{L}_\mathrm{bound}$ are 80 pixels, 5000 interior points, and 1000 free-stream points, respectively. Our boundary loss consists solely of a free-stream density condition, with no special treatment of the sphere-body surface. For each pixel in a batch, we sample 10,000 points within the corresponding frustum (or along a solitary ray when using the pinhole model). This number yields a reasonably-converged estimate of the volume (or line) integral \cite{Molnar2023b}. Specifically, our forward evaluations of Eq.~\eqref{equ:ray equation:cone} (or \eqref{equ:ray equation:thin}) have a standard deviation within 2\% of the long-run value. We also account for occlusions due to the test article by nullifying blocked rays.\par

\begin{figure*}[ht]
    \centering
    \includegraphics[width=6.5in]{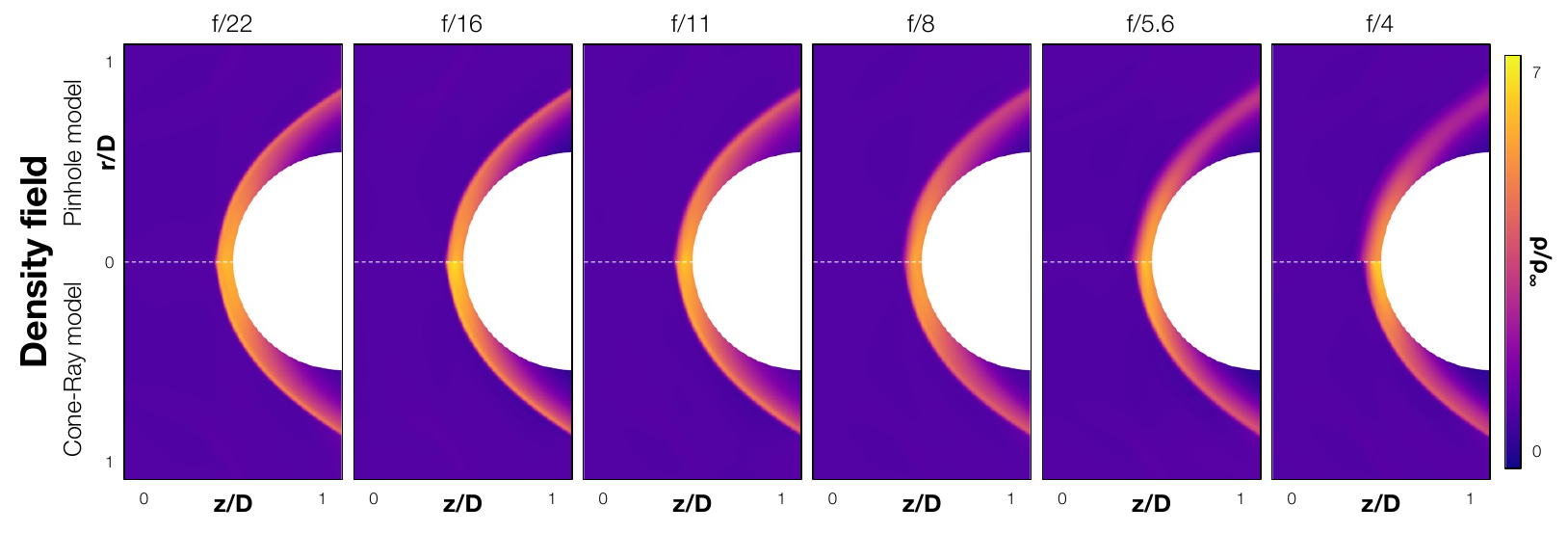}
    \caption{Reconstructions: (top) pinhole estimates and (bottom) cone-ray estimates for each $f$-number.}
    \label{fig:reconstructions}
\end{figure*}

Figure~\ref{fig:reconstructions} contains a panel of NIRT reconstructions. Estimates obtained with our cone-ray model, i.e., using Eq.~\eqref{equ:ray equation:cone}, are shown along the bottom row; pinhole estimates computed via Eq.~\eqref{equ:ray equation:thin} are plotted in the top row; and each column corresponds to the $f$-number listed in the header. Broadly speaking, all the density fields match our expectations about hypersonic flow over a spherical bluff body based on previous experimental results \cite{VanDyke1958, Schwartz1956}. All the cone-ray reconstructions exhibit a sharp spike in $\rho$ located a few millimeters upstream of the test article, with a maximum density next to the sphere at the axis of symmetry. Conversely, there is a stark degradation in the shock profile across the pinhole estimates with decreasing $f$-number. Unfortunately, the latter results are typical of reconstructions in the literature since the forward models used by existing algorithms for BOS tomography do not account for the camera's depth of field. The deleterious effects of such ``model errors'' are even visible in the $f/8$ and $f/11$ cases; these aperture settings are oft used in BOS tests, which underscores the need to use an accurate forward model in the reconstruction algorithm. Nevertheless, proper interpretation of the data with a (more) physical model (viz., conical rays) yields good results across the board.\par

\begin{figure*}[ht]
    \centering
    \includegraphics[width=6.5in]{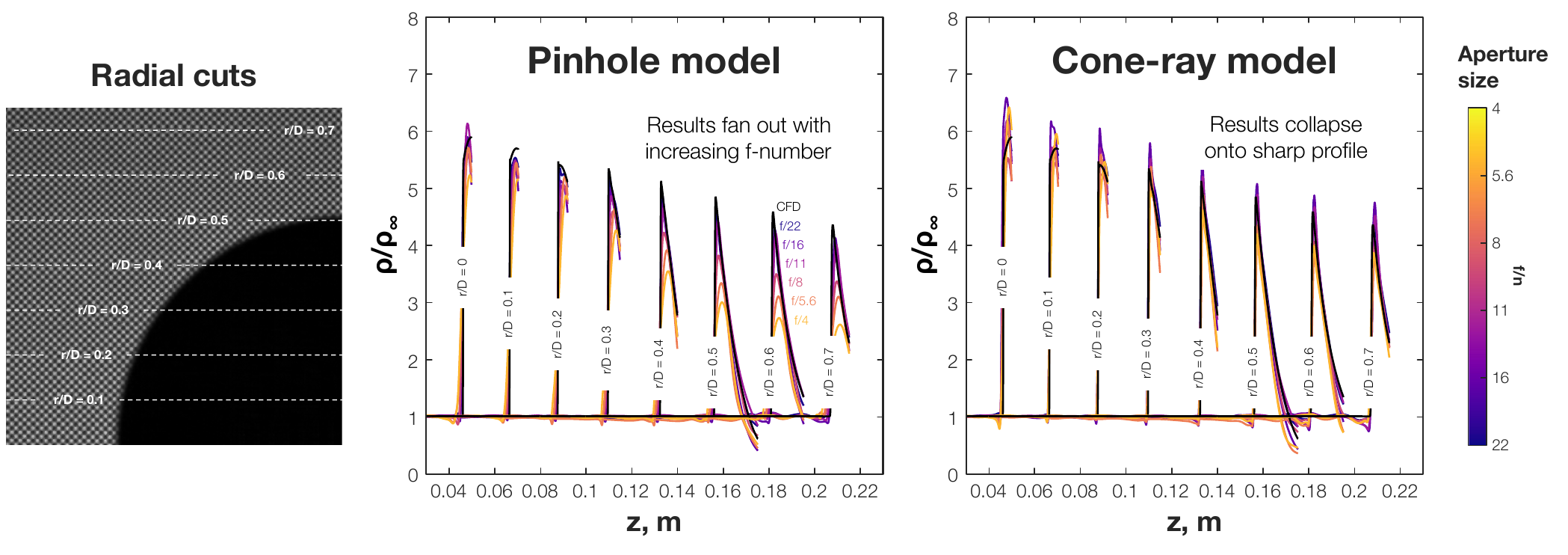}
    \vspace*{-0.75em}
    \caption{Axial distribution of $\rho$ at 0.1 $r/D$ increments. The pinhole profiles (left) smooth-out with larger apertures, while cone-ray reconstructions (right) collapse onto the CFD result.}
    \label{fig:cuts}
\end{figure*}

To better illustrate the effects of our cone- and thin-ray models during inversion, we examine cut plots of the density fields shown in Fig.~\ref{fig:reconstructions}. Figure~\ref{fig:cuts} presents 1D axial profiles of the reconstructed density ratio along the cuts indicated in the photo on the left-most side. In these plots, we color code the lines to indicate the $f$-number of the data set used to reconstruct the flow. For ease of viewing, we add a faux axial offset of 0.02~m with increasing $r/D$ to provide some horizontal spacing between the peaks. The central graph contains results based on the pinhole model, and the cone-ray results are shown in the right-most plot. Cuts from the CFD simulation are plotted in black in both cases. First, note that reconstructions based on the cone-ray model feature sharper density profiles than those computed with a pinhole model, even in the $f/22$ case. Second, note that the pinhole reconstructions decay and diffuse very quickly with the camera's $f$-number, broadening the shock profile. By comparison, the cone-ray results are mutually consistent, independent of the aperture setting. Third, we note that additional physics can be included in the inversion through the use of data assimilation, as demonstrated in \ref{app:data assimilation} (and omitted here to showcase the utility of our model, per se). Doing so has the potential to further enhance the accuracy of reconstructions, and it yields estimates of latent (unmeasured) fields like velocity and pressure.\par

We note that inversion of the cone-ray model becomes increasingly difficult with aperture size, leading to some artifacts that we discussed later in this section. We also acknowledge that, despite the marked improvement associated with our cone-ray model, the cone-ray reconstructions are not perfect. Indeed, acute discrepancies between these reconstructions are apparent in cuts close to the centerline. The reason for these errors is not known at this time.\par

\begin{figure*}[ht]
    \centering
    \includegraphics[width=5in]{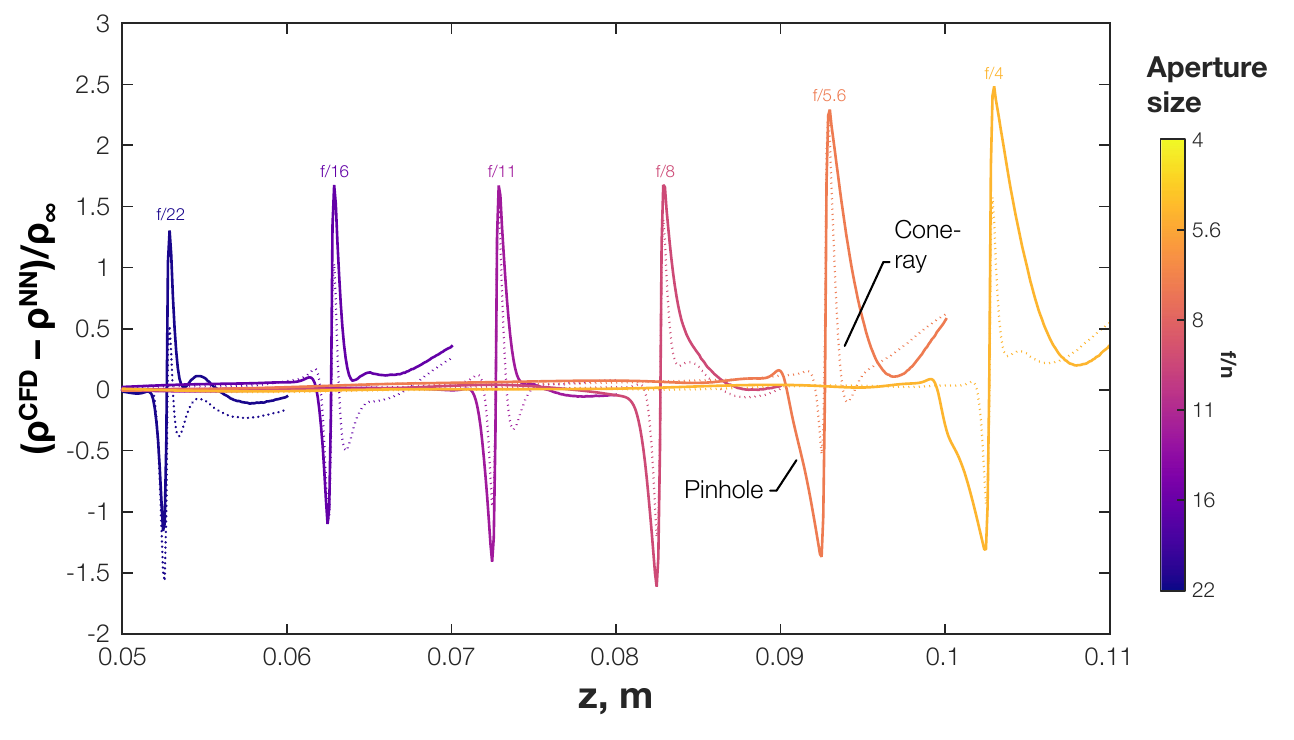}
    \caption{Normalized differences between the CFD- and BOS-based density fields from the $r/D = 0.4$ cut for each $f$-stop: the solid and dotted lines show cone-ray and pinhole results, respectively.}
    \label{fig:diff cuts}
\end{figure*}

In addition to the cut plots shown in Fig.~\ref{fig:cuts}, we also visualize discrepancies between our CFD data and reconstructions. Figure~\ref{fig:diff cuts} depicts differences between the CFD-based and BOS-tomography-based density ratios, using both the cone-ray (dashed line) and thin-ray (solid line) reconstructions for all six aperture settings. We visualize the $r/D = 0.4$ cut to avoid the issue of optical occlusion by the bluff body. Once again, an arbitrary axial offset is introduced to provide spacing between the peaks. Discrepancies between the CFD results and reconstructions increase with the aperture setting, but in each case, the cone-ray estimates are much closer to the CFD fields than the thin-ray estimates -- even for the $f/22$ data set, which most closely approximates pinhole imaging. The area under these difference curves is a metric that can be used to gauge the accuracy of reconstructions, assuming a reasonable CFD simulation. Table~\ref{tab:residuals} provides a summary of these areas, integrated over the reconstruction domain. As expected, when the aperture is sufficiently small ($f/22$ and $f/16$ in this experiment), discrepancies between the pinhole and cone-ray reconstructions are minimized. However, blurry artifacts become pronounced as the frustum grows larger. Notably, differences between the pinhole results and CFD simulation are 40\% and 70\% greater than cone-ray differences for the $f/5.6$ and $f/4$ cases, respectively. This result highlights the gains to be had from using a realistic cone-ray model for BOS tomography, as opposed to a pinhole model.\par

\begin{table*}[ht]
    \caption{Volume-Averaged Density Ratio Residuals}
    \centering
    \tabcolsep=0.2cm
    \begin{tabular}{c r r r r r r}
     \hline\hline
     \\[-2ex]
     \multirow{2}{*}{Model} & \multicolumn{6}{c}{Aperture Setting} \\ 
     & \multicolumn{1}{c}{$f$/22} & \multicolumn{1}{c}{$f$/16} & \multicolumn{1}{c}{$f$/11} & \multicolumn{1}{c}{$f$/8} & \multicolumn{1}{c}{$f$/5.6} & \multicolumn{1}{c}{$f$/4}\\
     \hline \\[-1.5ex]
     \multirow{ 1}{*}{}  Pinhole & $0.035$    & $0.047$ &   $0.040$ & $0.048$ & $0.088$ & $0.095$ \\
     \multirow{ 1}{*}{}  Cone & $0.035$    & $0.044$ &   $0.034$ & $0.032$ & $0.063$ & $0.056$ \\
     \\[-2ex]
     \hline\hline
    \end{tabular}
    \label{tab:residuals}
\end{table*}

Readers should note that all the results presented in this section are computed via UBOS. Using deflection data instead of the raw images -- via Eq.~\eqref{equ:OF data loss} instead of \eqref{equ:UBOS data loss} -- yields broadly similar results, but the deflection-based estimates exhibit larger free-stream artifacts. These artifacts are likely attributable to errors in the deflection data set, since UBOS operates directly on the images and does not require multiple stages of ad hoc regularization \cite{Grauer2020}. UBOS should be valid for the sphere flow data sets because the Settles number in these tests peaks at 0.27, which satisfies the $Se \ll 1$ criterion proposed in Sec.~\ref{sec:modeling:forward}. In previous work, we found it difficult to counteract noisy measurements by adjusting the relative weight of the $\mathcal{L}_\mathrm{meas}$ and $\mathcal{L}_\mathrm{phys}$ losses \cite{Molnar2022}. Instead, errors in the deflection data must be modeled and discounted, accordingly; in principle, this could be done using the joint-PDF of deflections.\par

Lastly, it bears mentioning that processing BOS data with a cone-ray algorithm becomes more challenging and less numerically stable as the $f$-number decreases. The action of the aperture is analogous to that of a blur kernel: increasing the width of the kernel makes the data smoother and more ambiguous with respect to changes in the underlying signal. Consequently, small perturbations to the data can imply a large change in the signal, resulting in the amplification of noise in the inversion, and this problem grows worse with the aperture size \cite{Daun2016}. Hence, despite the utility of our cone-ray technique, there remains a trade-off between opening the aperture to increase the signal level versus keeping it closed to mitigate ambiguities in the deflection data.\par

\section{Conclusions and Outlook}
\label{sec:conclusions}
This work presents a novel cone-ray model for BOS tomography that accounts for depth-of-field effects, which limit the fidelity of BOS when using a conventional tomographic reconstruction algorithm. The developed approach stands in contrast to the ubiquitous pinhole model, which has been used in all prior examples of BOS tomography, to the best of our knowledge. We conduct forward and inverse analysis of BOS imaging using numerical and experimental data sets. Our experiment features hypersonic flow over a spherical bluff body, giving rise to a bow shock that exhibits very sharp density gradients. This shock enables us to assess the accuracy of our reconstructions by gauging the degree of smoothing along the shock front. To demonstrate our method, we repeated our sphere flow tests with six aperture settings and reconstructed all six data sets with a pinhole model as well as our new method. Pinhole estimates exhibited visible errors that worsened with decreasing $f$-numbers: damping and broadening the shock profile. By contrast, results computed with our cone-ray model were sharply peaked and mutually consistent, lending credence to our approach.\par

Two important conclusions can be drawn from this work:
\begin{enumerate}
    \item Blur is significant in BOS imaging scenarios with large density gradients, leading to meaningful differences between the deflections predicted with a pinhole model and those corresponding to a cone-ray model. Experimental results in this paper demonstrate the superiority of cone-ray reconstructions over pinhole reconstructions. 
    
    \item Reconstructions are susceptible to errors that stem from the deflection sensing step. Therefore, to minimize artifacts and recover sharp flow structures, we recommend tailoring the setup to stick within the linear regime of optical flow ($Se \ll 1$) and processing raw image data with a UBOS algorithm.
\end{enumerate}\par

\appendix
\renewcommand{\thesection}{Appendix \Alph{section}}

\section{Neural Architecture}
\label{app:network}
The network $\mathcal{N}$ comprises an input layer, output layer, and series of $N_\mathrm{l}$ hidden layers,
\begin{subequations}
    \label{equ:network}
    \begin{align}
        \mathcal{N}\mathopen{}\left(\mathbf{z}^0\right) &= \mathbf{W}^{N_\mathrm{l}+1}\left[\mathcal{N}^{N_\mathrm{l}} \circ \mathcal{N}^{N_\mathrm{l}-1} \circ \dots \circ \mathcal{N}^2 \circ \mathcal{N}^1\mathopen{}\left(\mathbf{z}^0\right)\right]+\mathbf{b}^{N_\mathrm{l}+1},
        \intertext{with}
        \mathbf{z}^l = \mathcal{N}^l\mathopen{}\left(\mathbf{z}^{l-1}\right) &= \sigma\mathopen{}\left(\mathbf{W}^l\mathbf{z}^{l-1} + \mathbf{b}^l\right) \quad\text{for}\quad l \in \{1, 2, \dots, N_\mathrm{l}\}.
    \end{align}
\end{subequations}
The vector $\mathbf{z}^l$ contains the value of neurons in the $l$th layer, $\mathbf{W}^l$ and $\mathbf{b}^l$ are the weight matrix and bias vector for the $l$th layer, and $\sigma$ is a non-linear activation function that is applied to each element of the argument. In this work, the input vector $\mathbf{z}^0$ is simply $\mathbf{x}$. All the trainable weights and biases in $\mathcal{N}$ are included in $\boldsymbol\uptheta$. We use swish activation functions,
\begin{equation}
    \label{equ:method:swish}
    \sigma(z) = \frac{z \,\exp(z)}{1 + \exp(z)},
\end{equation}
which have been shown to improve the stability of gradient flow in training compared to hyperbolic tangent functions, among others \cite{Molnar2022}.\par

\section{Physics-Informed BOS}
\label{app:data assimilation}
This appendix reports a data assimilation procedure, a.k.a. ``physics-informed BOS,'' applied to the hypersonic sphere flow case. Instead of a heuristic penalty like the TV loss, NIRT algorithms can be augmented with physics-based equations. Hypersonic flow over a spherical bluff body is approximately governed by the steady, axisymmetric, compressible, and inviscid continuity, momentum, and energy equations:
\begin{subequations}
    \label{equ:Euler}
    \begin{align}
        e_\mathrm{c} &= \nabla \cdot \left(\rho \mathbf{v} \right), \\ 
        \mathbf{e}_\mathrm{m} &= \nabla \cdot \left(\rho \mathbf{v}\mathbf{v}^\top \right) + \nabla p,~\text{and} \\
        e_\mathrm{e} &= \nabla \cdot \left[\left(\rho E + p \right) \mathbf{v}^\top\right],
    \end{align}
\end{subequations}
respectively, where $\mathbf{v}$ contains the axial and radial components of velocity, $p$ is pressure, $E$ is the total energy of the fluid, and the del operator is defined for a cylindrical system of coordinates. All physical terms in Eq.~\eqref{equ:Euler} have been placed on the right side of the equalities, leaving a scalar or vector residual on the left side, denoted $e$ or $\mathbf{e}$. We also note that Eq.~\eqref{equ:Euler} contains four equations and five unknowns and must be closed with an equation of state, e.g.,
\begin{equation}
    p = \left(\gamma - 1\right)\rho\left(E - \frac{1}{2} \mathbf{v} \cdot \mathbf{v}\right),
    \label{equ:state}
\end{equation}
wherein $\gamma$ is the ratio of specific heats.\par

In physics-informed BOS \cite{Molnar2023a}, the NIRT is modified to use these equations for regularization instead of an ad hoc penalty term. Accordingly, the volumetric integral contains the residuals from Eq.~\eqref{equ:Euler},
\begin{equation}
    \mathcal{L}_\mathrm{penalty} = \frac{1}{\left|\mathcal{V}\right|} \iiint_\mathcal{V} e_\mathrm{c}^2 + \left\lVert \mathbf{e}_\mathrm{m} \right\rVert^2  + e_\mathrm{e}^2 \,\mathrm{d}\mathcal{V}. 
    \label{equ:physics}
\end{equation}
Only minimal changes to the NIRT are needed to perform a physics-informed reconstruction. The most notable is a change to the mapping,
\begin{equation}
    \mathcal{N}: \mathbf{x} \mapsto \left(\rho, \mathbf{v}, E\right).
\end{equation}
Free-stream conditions for the additional fields are incorporated into $\mathcal{L}_\mathrm{bound}$, similar to the treatment of $\rho$ in Eq.~\eqref{equ:boundary}. Several hard constraints are implemented to improve the stability of training, as reported in \cite{Molnar2023a}. Another modification that we find helpful, which was not included in \cite{Molnar2023a}, is a local compressibility weighting factor that modulates the relative influence of penalties throughout the domain,
\begin{equation}
    \frac{\mathrm{max}\mathopen{}\left( \left\lVert \nabla \rho \right\rVert \right)}{\left\lVert \nabla \rho \right\rVert + \mathrm{max}\mathopen{}\left( \left\lVert \nabla \rho \right\rVert \right)}, \nonumber
\end{equation}
where $\mathrm{max}(\cdot)$ returns the batch maximum for each evaluation of Eq.~\eqref{equ:physics}. This term scales the integrand in Eq.~\eqref{equ:physics}. Besides these changes, optimization proceeds as described in Sec.~\ref{sec:modeling:inverse}.\par

\begin{figure*}[ht]
    \centering
    \includegraphics[width=5.5in]{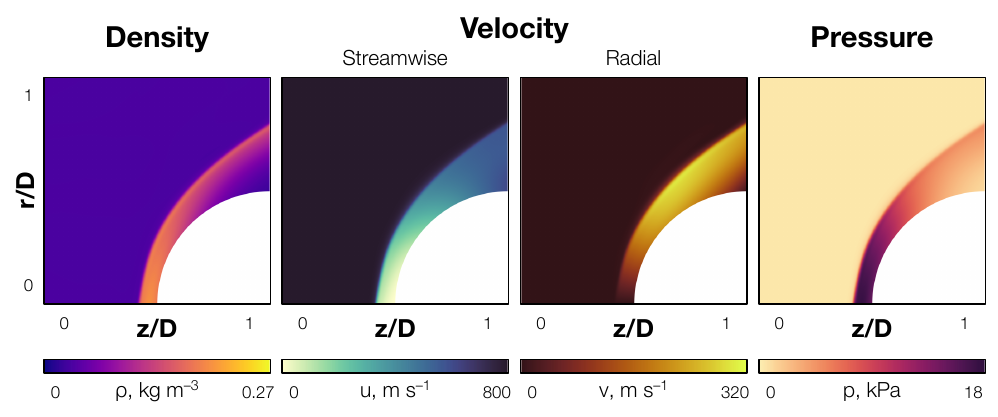}
    \vspace*{-0.75em}
    \caption{Physics-informed estimates of density, velocity, and pressure based on the $f/22$ data set.}
    \label{fig:data assimilation}
\end{figure*}

Finally, results of the data assimilation procedure are shown in Fig.~\ref{fig:data assimilation}. These estimates of density, velocity, and pressure are obtained from the $f/22$ images via physics-informed BOS without using a conventional CFD solver to train the network. Moreover, the fields in Fig.~\ref{fig:data assimilation} are phenomenologically consistent with our prior knowledge of hypersonic flow over a sphere as well as experimental measurements thereof \cite{VanDyke1958}. In this case, the algorithm seeks a flow state that is both consistent with the measurements and satisfies the governing equations (i.e., the compressible Euler equations), which is akin to ``inverse CFD.'' Data from additional diagnostics, like pressure taps, thermocouples, infrared thermography, and laser diagnostics, can be included in the inversion through supplemental data loss terms. The framework provides a powerful tool for synthesizing on- and off-body measurements with known physics equations to paint a rich picture of the flow dynamics.\par

\section*{Acknowledgments}
This material is based upon work supported by NSF under grant no.~2227763, NASA under grant no. 80NSSC19M0194, FAU Erlangen-N{\"u}rnberg, and a DoD NDSEG Fellowship. The views and conclusions contained herein are those of the authors and should not be interpreted as representing the official policies or endorsements, either expressed or implied, of NSF, NASA, DoD, or the U.S. Government.\par

The authors also gratefully acknowledge K. Kim and other members of the Combs Research Group at UTSA for helping with the BOS setup and operating the wind tunnel facility.\par

\bibliographystyle{osajnl2} 

\end{document}